# Vision to Reality: From Robert R. Wilson's Frontier to Leon Lederman's Fermilab

Lillian Hoddeson and Adrienne Kolb[*]

## Abstract

This paper examines the roles of vision and leadership in creating and directing Fermi National Accelerator Laboratory from the late 1960s through the 1980s. The story divides into two administrations having different problems and accomplishments, that of Robert R. Wilson (1967-1978), which saw the transformation from cornfield to frontier physics facility, and that of Leon Max Lederman (1979-1989), in which the laboratory evolved into one of the world's major high-energy facilities. Lederman's pragmatic vision of a user-based experimental community helped him to convert the pioneering facility that Wilson had built frugally into a laboratory with a stable scientific, cultural, and funding environment.

## Introduction

In the fall of 1978, when Leon Max Lederman stepped into Robert Rathbun Wilson's shoes to become the second director of Fermi National Accelerator Laboratory, he found them rather large. He could not think of walking in them comfortably. "I'm a good physicist," Lederman conceded, "but Bob is a great man."[1]

In 1967, Wilson had set out to design and build a facility of unprecedented scope and cost on the Illinois prairie. Its site was in Batavia, about thirty miles west of Chicago.[2] By 1972, he had completed the accelerator known as the Main Ring, with an energy of 500 BeV (billion electron volts), more than double the agreed upon 200 BeV. Moreover, drawing on his bold imagination and relentless frugality, he delivered the machine ahead

---

[*] Lillian Hoddeson is Professor of History at the University of Illinois in Urbana-Champaign and Historian at Fermilab in Batavia, Illinois. She has written extensively on twentieth-century solid-state physics and big science. Her most recent book, with Michael Riordan, is *Crystal Fire: the Birth of the Information Age* (N. Y.: W. W. Norton, 1997). Presently she is completing, with Vicki Daitch, *True Genius: the Life and Science of John Bardeen*, to be published in 2002 by Joseph Henry Press, and coauthoring with Michael Riordan, Adrienne Kolb, and Steve Weiss, the forthcoming *Tunnel Visions: the Rise and Fall of the Superconducting Super Collider*.

Adrienne Kolb is Archivist at Fermilab in Batavia, Illinois. She is coauthor of: "The Superconducting Super Collider's Frontier Outpost, 1983-1988," *Minerva* (2000); "The Mirage of the World Accelerator for World Peace and the Origins of the SSC," *Historical Studies in the Physical and Biological Sciences* (1993); "A New Frontier in the Chicago Suburbs: Settling Fermi National Accelerator Laboratory, 1963-1972," *Illinois Historical Journal*, (1995); and the forthcoming *Tunnel Visions: The Rise and Fall of the Superconducting Super Collider*.

of schedule, under budget, and with three times the promised number of experimental areas.[3]

Wilson's difficulties began when he then immediately proceeded to apply the unspent six million dollars allocated for the Main Ring toward a scheme for doubling its energy by circulating its proton beam through an added ring of superconducting magnets. This technology had not yet been tried for a large proton synchrotron. Wilson called the new ring the Energy Doubler; the machine was renamed the Energy Saver during the nation's energy crisis.

The United States Department of Energy (DOE) did not approve Wilson's unofficial reallocation of funds. He was required to return the excess to the federal treasury. Construction of the Doubler was then stalled by heated controversy between Wilson and the DOE over future funding.[4] For this reason, Lederman "found good and bad news" on taking the reins in 1978. Among the good news was that the "the group that Bob Wilson had assembled was really magnificent." Moreover, "the vision of a superconducting accelerator (Energy Doubler/Saver) was very clear. The tradition of architecture and ecology were firmly established, as was the style which somehow strove to preserve individuality within the need for impressive collaborations." The bad news was that "in the transition, the focus had been lost, the Lab was impoverished, and the physics program was hostage to an under-funded Saver project: a project whose physics goals were ill-defined."[5]

Having worked for thirty years at the front lines of particle physics, even in Wilson's leaky trenches, Lederman offered what Fermilab needed to fulfill its promise as the world's highest energy physics laboratory. Drawing on his rhetorical skill, sense of humor, ability to negotiate, and his boundless passion for physics, Lederman made allies of Wilson's antagonists. He upgraded facilities that Wilson had built frugally into long-term structures. He made Fermilab a stable environment for physics research.

This paper examines the roles of vision and leadership in creating and directing a new physics instrument in the midwest from the late 1960s through the 1980s. The story divides into two administrations having different problems and accomplishments. The first, under Wilson, saw the transformation from cornfield to frontier physics facility; the second, under Lederman, witnessed the evolution of the initial facility into one of the world's major scientific laboratories.

## Wilson's Vision

The vision with which Wilson built Fermilab reflected his own self-image, one he often expressed in terms of three figures: the pioneer, the craftsman (or blacksmith), and the Renaissance man. The pioneer was for Wilson an explorer who boldly and independently confronts the limits of known territory, including man's understanding of Nature. "Research plays a present-day role analogous to the role that opening of the west played at an earlier stage in our country," he wrote in the period when he built Fermilab.[6]

Unlike the brutal conquerers described by historians of the American West, Wilson's pioneer was a noble frontiersman whose explorative ventures were shaped by democratic ideals. The image matched the self-sufficient explorer described in the 1890s by the historian Frederick Jackson Turner.[7] Wilson applauded the pioneer's can-do attitude, his resourcefulness and his willingness to cope practically with fundamental matters. Since 1945, when Vannevar Bush wrote *Science, the Endless Frontier*, many physicists had associated the search for new scientific information with exploring frontiers.[8] Frontier imagery had also appealed to a number of the physicists whom Wilson admired, including his Berkeley mentor, Ernest O.Lawrence, and his Berkeley colleague and director at Los Alamos, J. Robert Oppenheimer.[9]

Wilson chose for his laboratory an institutional model drastically different from the scholarly university selected by Stanford Linear Accelerator Center (SLAC), or the rugged military camp chosen by Brookhaven National Laboratory (BNL).[10] His model was a peaceful prairie, with a large expanse of open space, tall prairie grass, and natural habitats for native wildlife, even a herd of bison.

For this frontier site, he designed a spare, but powerful accelerator where physicists would conduct simple, elegant, and clever experiments. The accelerator beam was to circulate in an underground tunnel, with little visible disruption on the surface. He envisioned a machine of "great beauty" that would "add to the satisfaction of our lives." The project would be "primarily spiritual."[11]

As for the laboratory community, he envisioned a utopian "science city" governed by noble ideals: democracy, civil rights, equality, respect for diversity and environmental preservation.[12] "In the course of giving a very large acceleration to our particles, let us hope that we can contribute at least a small acceleration to society," Wilson said.[13] His response to questioning from Senator John Pastore in April 1969 about the role of the lab regarding national defense has become legend: "It has to do with, are we good painters, good sculptors, great poets? I mean all the things we really venerate and honor in our country and are patriotic about. It has nothing to do directly with defending our country except to make it worth defending."[14]

Wilson's second image, the hands-on craftsman, was no less important to his vision. When machinery broke down during Wilson's boyhood, he would avoid the long and lonesome ride into town by visiting the blacksmith's shop, where he would hammer out a replacement part. "Working with him, I had the confidence that with your own hands you could build large contraptions and make them work. Then I used to work in the blacksmith shop just for the hell of it, and learned how to use my hands and make things. I think that was a very useful part of my training."[15] Like the blacksmith, the engineers at Edison's Menlo Park, or the physicists working on the Manhattan Project, Wilson was a pragmatist. In the tradition of William James and John Dewey, Wilson trusted hands-on experience.[16]

Wilson's most comprehensive image of himself was expressed by his third figure, the Renaissance man, in whom art, spirit, and nature are unified. Some historians have looked toward "Renaissance naturalism" to explain the birth of modern science.[17] In Wilson such a naturalism awoke about 1954, when he was in Paris studying art. The sensibility came into focus about 1965 when he was again studying art and sculpture in Europe. He compared Gothic cathedrals with accelerators, finding that both express the spirituality of their age. He felt that both achieved their aspirations and expression when form, function, and focus worked together in an upward surge of purpose.

While thinking in these terms, Wilson received a copy of the 1965 proposal put forth by a Berkeley team for a 200-BeV proton synchrotron. Wilson criticized the design. He considered it too expensive and overdesigned. It fell short of the heights of creativity he expected of a large accelerator. Wilson's aesthetics called for frugality in building apparatus and facilities. On artistic grounds he believed "that something that works right away is overdesigned and consequently will have taken too long to build and will have cost too much." He worried that the "exorbitant" expense of accelerators would halt the advance to higher energy. A component built frugally might not immediately work as it should, but it could be fixed, modified, or even recycled[18] This had been the attitude at Berkeley in earlier times, in particular, when Wilson studied there under Lawrence in the 1930s. Wilson made his criticisms of the proposal known.[19]

Less than two years later, Wilson was offered the chance to build the new several hundred GeV accelerator his way. In December 1966, President Lyndon Johnson and the Atomic Energy Commission (AEC) approved for the project a 6800-acre site in Weston, Illinois. The open spaces and skies of this prairie expanse became a blank canvas upon which Wilson would design the laboratory that became Fermilab. We have elsewhere described how Wilson, his Associate Director Edwin Goldwasser, and many others, built Fermilab to fit his Renaissance aesthetics and ideals.[20] Suffice it to say he was granted considerable decision-making power, both by the AEC and by the Universities Research Association (URA), the university consortium created to govern the new laboratory.[21]

Unlike Brookhaven and Berkeley, criticized for favoring users from their own geographical regions, the new laboratory would welcome proposals from all regions of the United States. A national laboratory in which users would be "at home and loved" had been a theme of an earlier paper by Leon Lederman, then a user at Brookhaven.[22] It had disturbed Lederman when Brookhaven granted beamtime prefentially to its east-coast users and when young physicists had trouble getting in the door. At a Brookhaven users' meeting in 1963, Lederman proposed the notion of the "Truly National Laboratory," or TNL. Unlike BNL, the TNL would be a users' paradise offering complete on-site facilities for outside users, scheduling and advisory committees to assure fairness in the allocation of beamtime, and free communication between management and users. The accessibility of the facilities would be "*a right* to any physicist bearing a competitively acceptable proposal."[23]

Wilson and Goldwasser tried to realize this vision of the TNL at the new midwest facility. To emphasize the centrality of its role, Wilson secured the consent of President Johnson to name the laboratory "*The* National Accelerator Laboratory," or NAL.[24] Users from around the world would in fact dominate the research program of the laboratory. It was specified that 75% of the research program would be performed by users and 25% by resident staff members.[25] One of Wilson's reasons for appointing Goldwasser as his Associate Director was his prior success in instituting a comparable user policy at Argonne National Laboratory. NAL was later renamed Fermi National Accelerator Laboratory, or Fermilab, in 1974.

Wilson found it difficult to attract permanent staff to NAL. He drew on rhetoric as a tool. As a boy growing up on the plains of Wyoming, he had listened to his male relatives spin tales around the campfire and learned that "it was how you told the story that mattered."[26] His recruiting efforts appealed to the sense of community that binds compatriots on a common venture. "Most of all, I want someone to come and say, `I'm going to work on this job, and I am going to live in this area, and this is the laboratory with which I'm going to identify, and I am going to be committed to its success."[27] Often Wilson's rhetoric drew on frontier narratives. In 1973, addressing the difficulties of the primitive working conditions at the laboratory, he told experimentalists that they were "presently ready to embark on [an] adventure," to unravel Nature's mysteries, which at this juncture were "just as much of a challenge to the experimenter as they were when our pioneer forebears started at the beginning of the century."[28]

Wilson's vision was initially compelling to his staff and to the Joint Committee on Atomic Energy (JCAE) in Washington D.C. But over time the picture revealed inconsistencies in his philosophy, for instance, between his frugality and goal to achieve "higher energy or bust," or with his commitment to users. Wilson agreed with Lederman's philosophy of a "user's paradise,"[29] but in the managerial structure of his laboratory, he was careful to draw distinctions between the roles of users and competing interests represented by the members (supposedly unbiased) of the laboratory's external Program Advisory Committee, as well as his own URA Board of Trustees. Moreover, Wilson's policy of providing minimal experimental equipment conflicted with the trend to conduct much larger and more precise experiments. Users did not agree that "providing expensive facilities, even any facility, may tend to paralyze better developments later on." Nor did they accept that "techniques and research interests change so rapidly that it is better to let the major part of these areas grow out of the actual ideas and demands of the experimenters at the time they use the machine."[30] Experimenters did not appreciate the functional flaws of the architecturally interesting experimental halls built under Wilson. They complained about working in leaky tunnels, old barns, and other inadequate structures, while Wilson kept his eyes on the future, directing much of the lab's funding to the Energy Doubler.

It also became clear that the very notion of a collaborative venture of lone frontiersmen is paradoxical. For, as in any effort to herd cats, there is an intrinsic paradox in demanding independent users to stay in line with the dictates of a strong director.[31] Other inconsistencies of Wilson's vision concern the long life of a laboratory created almost entirely of makeshift structures.  On the level of patronage, Wilson's conviction that physicists should have privileged status in requests for government funding was revealed as a relic of the postwar years.  In the late 1970s, politicians and their constituents no longer related to the social value of the work of physicists, notwithstanding Wilson's noble words to Senator Pastore in 1969.

Wilson was among the first to recognize certain contradictions in his vision.  In an essay published in *Daedalus* in 1970, "My Fight Against Team Research," he explained that "as a young man" he had sought the life of "the lone scientist in pursuit of truth."  He had "accepted the cliche" that individual research was "creative, poetic, and enduring," while team research was "superficial, uncreative and dull."  But when he came to study the nucleus he discovered that "it is almost as hard to reach the nucleus by oneself as it is to get to the moon by oneself."  He embarked on team research, but retained "ambivalence" and "prejudice" against the notion.[32]

Wilson's leadership ultimately felt the strain of inadequate funding.  In his early years as director, he had dealt easily and informally with the Atomic Energy Commision (AEC), which had emerged out of the Manhattan Engineer District.  His chain of Los Alamos and Berkeley connections, indeed gave him an insider's conduit up to his old friend Glenn Seaborg and the Joint Congressional Committee on Atomic Energy.  But during Wilson's tenure, the funding environment changed suddenly and dramatically when, in January 1975, President Gerald Ford's administration transformed the AEC into two new agencies:  the Energy Research and Development Administration (ERDA) and the Nuclear Regulatory Commission (NRC).  ERDA and the Department of Energy (DOE), which replaced ERDA in October 1977 during President Jimmy Carter's administration, were different; unlike the AEC they were organized to deal with *all* the nation's energy needs, not only with atomic and basic energy research.  Funding for high-energy physics research was now in competition with many diverse public projects.  Like the leaders of the other projects, the directors of national laboratories were expected to travel to the Capital to justify their funding requests.  Wilson objected.  His job, he felt, was to lead his people on his site.[33]

The conflict between Wilson and Washington reached crisis proportions after John Deutch came into office in 1977 as Director of the Office of Energy Research.  Deutch questioned Wilson's assumption that DOE was committed to fund the Doubler.  The conflict stalled negotiations and the Doubler sat in limbo.  Wilson resigned in protest early in 1978.  In May URA accepted Wilson's resignation.

Then URA proceeded to seek Wilson's successor.  The first offer went to SLAC's Burton Richter, a 1976 Nobel Prize winner.  He declined.  In mid-summer 1978, URA nominated Lederman.  As Fermilab's highest profile user, and with an outstanding record of research achievement and scientific statesmanship, Lederman was a natural candidate.  Wilson was among those who strongly encouraged Lederman to take the job.  Lederman agreed to serve as Director Designate until June 1, 1979, and then assume full office as Director.  In the interim, Philip V. Livdahl continued to serve as Acting Director.[34]

**Lederman's Vision**

Fermilab's guiding vision switched with the change in director. Like Wilson, Lederman structured his vision on his self-image. It was unabashedly romantic: it rested on his all-embracing passion for physics research. But it was practical, too.

Lederman often admitted in his public talks and popular writings that "the life of a physicist is filled with anxiety, pain, hardship, tension, attacks of hopelessness, depression, and discouragement." But he insisted that physics was worth the pains, because the study offered two special pleasures. One was to experience rare "Eureka" moments of discovery -- "punctuated by flashes of exhilaration, laughter, joy, and exultation." These "epiphanies" derive from "the sudden understanding of something new and important, something beautiful."[35] He explained that "the best discoveries always seem to be made in the small hours of the morning, when most people are asleep, when there are no disturbances and the mind becomes contemplative." The experimenter sits off in a lonely laboratory staring at numbers. "You look and look, and suddenly you see some numbers that aren't like the rest -- a spike in the data. You apply some statistical tests and look for errors, but no matter what you do, the spike's still there. It's real. You've found something. There's just no feeling like it in the world."[36]

Almost as exciting were the everyday tasks of research. In one anecdote from the 1950s, Lederman and Gilberto Bernardini, a visiting experimentalist from Rome, build a particle counter at Columbia University. Working late into the night, long after the machinists had gone home, they finished insulating, soldering, and flushing gas through the charged system to which an oscilloscope was attached. Suddenly Bernardini went "stark, raving wild" and cried "Mamma mia! Regardo incredibilo! Primo secourso!" Shouting and pointing, he lifted Lederman up in the air and danced him around the room. "What happened?" Lederman asked. "Izza counting. Izza counting!" Lederman joined Bernardini in the excitement over having built "with our hands, eyes, and brains" a device that could detect cosmic-ray particles.[37]

Lederman felt that a national physics laboratory should be a total environment for nurturing both the epiphanies and the day-to-day work of experimental research in one "user's paradise." While on its basic philosophical level Lederman's vision hardly differed from Wilson's, the details were different. Realizing Lederman's idea meant addressing the needs of users. Having spent most of the last three decades immersed in experimental research at numerous laboratories in the United States and Europe, Lederman came into his position as Fermilab director with the authority to promote his user-friendly vision for the laboratory. He was widely known for his good taste in selecting fertile physics problems and for his many successful experiments.[+]

---

[+] They included, the discovery of the long-lived neutral K meson; the magnetic moment of the muon using magnetic resonance; parity violation in the muon's beta-decay, the "two-neutrino experiment" demonstrating the neutrinos that arise from muon decay (for which he shared the 1988 Nobel Prize with Melvin Schwartz and Jack Steinberger); evidence for the particle later known as the J/psi (for which Samuel Ting and Burton Richter won the 1976 Nobel Prize); and the discovery in 1977 of a family of Upsilon particles, the first evidence for the bottom quark.

*The Immediate Crisis of the Doubler and the Call for Outside Advice*

Lederman's term as Director Designate began in crisis. The accelerator builders and experimenters at the laboratory could not wait until he became Director to know whether they would compete with CERN in the race to discover the W and Z vector bosons. In 1975 and 1976, Carlo Rubbia had proposed that Fermilab convert its Main Ring into a colliding-beams machine to achieve the higher energies needed to detect these particles, but Wilson and his Physics Advisory Committee rejected the idea. [38] Rubbia then took the proposal to CERN, where it was accepted.

Fermilab could compete only if it built a collider of comparable energy based on the Main Ring. That however implied giving up Wilson's fight for the Doubler, at least in the foreseeable future. An important question was whether the Doubler was in fact feasible. To investigate, Lederman called on experts. He appointed "three wise men" to serve as the Doubler Review Committee: Matthew Sands of Caltech, Richter of the Stanford Linear Accelerator Center (SLAC), and Boyce McDaniel of Cornell. Discussing Fermilab's technical issues from October 1978 to January 1979, this committee gave Lederman confidence in the Doubler's technical viability.

To address a range of questions about Fermilab's future, Lederman issued an invitation to all of the Fermilab physicists to an Armistice Day "shootout."[39] Meeting all day and into the night on November 11, 1978, the group sorted out the possibilities. After eighteen hours of heated discussion, Lederman had his decision. Fermilab would forego the race for the W and Z and would complete the Doubler, so that in the future the laboratory could collide beams at higher energy and preserve its front-line position at the energy frontier.[40]

Completed in 1983, the Doubler became the basis for Fermilab's 800 GeV accelerator, the Tevatron.[41] The laboratory's new initiative was divided into two programs: a proton-antiproton colliding-beams program with energies close to 2 TeV (trillion electron volts) known as Tevatron I; and the enhanced fixed-target program operating at almost 1 TeV known as Tevatron II.

*Working with Washington*

Lederman had never before dealt with budgets as large as those of Fermilab, but he had had considerable experience working on issues of science funding. He had often served on national and international physics advisory panels, especially after 1961 when he was appointed Director of Columbia University's Nevis Cyclotron Laboratory. He was a founding member of the High Energy Physics Advisory Panel (HEPAP). This experience, together with his years of writing grant proposals, had brought him into an extensive network through which he quickly learned of available funding.

A gifted comedian, adept with words, he proved effective in the Washington corridors of power. Calmly assessing each audience, he seemed to know intuitively when to draw on science, emotion, or humor. Unlike Wilson, he resigned himself to his role in Washington. "I'm constantly shuttling to Washington to cajole federal and congressional officials into keeping budget cuts to a minimum," Lederman told an interviewer.[42]

Lederman's initial hurdle was the bitter legacy of Wilson's battles with Deutch over the Doubler. A circumstantial problem was DOE's heavy investment in ISABELLE, Brookhaven's rival superconducting-magnet project.[43] Lederman worked through and around the problems with the skill of a poker player. Fermilab's Bruce Chrisman vividly recalled Lederman's cool performance during one Washington visit in the early days of his tenure as director. The purpose of the trip was to review the finances of the accelerator division and to secure several million dollars promised for the Doubler.

In one scene, the Fermilab contingent sat tensely around a conference table in DOE's Forrestal building as they and several DOE officials waited for Deutch, who was late. When Deutch arrived, he sat down at the head, leaned back, and stretched his legs out across the table, as if displaying his authority. The subsequent interplay was "not exactly friendly," Chrisman reminisced. Deutch frequently interrupted Lederman's prepared remarks. Lederman ignored the jibes and continued his presentation. Other physicists might have lost their composure, but Lederman stayed calm and "didn't rise to the bait." Afterwards, Deutch released the funds.[44]

For his partner in dealing with the DOE, Lederman made effective use of Andrew Mravca, who was then working at Fermilab as DOE's in-house liaison. The two "new boys on the block" worked closely, informing the DOE of all progress at Fermilab and of the lab's compliance with DOE's bureaucratic requirements, for instance, the imposed "project management plan." Where Wilson had flatly rejected such procedures, Lederman and Mravka accomodated.[45] Fermilab's relationship with Washington improved dramatically. "The Wilson-Deutch problem was an unneccessary tragedy since they were both charismatic personalities who could have gotten along very well," Lederman later commented. "Each was stubborn and insisted on his prerogatives."[46]

### Building for Permanence

Lederman broke with Wilson's philosophy of building for the short term, even if that occasionally meant overspending. He recognized that inferior facilities were a liability for physicists competing on the world stage. Europe was now competitive with the U.S. in the field of high-energy physics.

Lederman added many research capabilities to Wilson's design, including sounder buildings and laboratories. The additions included a magnet factory, a muon experiment building, a new computing center, an antiproton source, and collision halls to house two mammoth and supremely complex particle detectors -- a magnetic detector, CDF (the Colliding Detector at Fermilab), and a rival nonmagnetic detector project, DZero. The collision facilities where hundreds of researchers collaborated in the same experiment allowed Fermilab to leap forward in its transition from big science to megascience.[47]

Lederman endorsed a substantial upgrade of the Tevatron, the Main Injector, which was completed during the administration of Fermilab's third director, John Peoples. This upgrade not only provided an increase of energy to 900 GeV but increased the luminosity of the proton-antiproton collider and doubled the fixed-target intensity. In his last years as director, Lederman also planned what he saw as Fermilab's next machine, the Superconducting SuperCollider (SSC). But his vision of building the SSC at Fermilab became clouded by many factors which we discuss in another work.[48]

*Realizing the Users' Paradise*

Lederman sympathized with the experimentalists working in Wilson's leaky pits. After all, he had been one of them. As Head of Fermilab's User's Organization, he had dedicated the laboratory in 1974 as the realization of his 'Truly National Laboratory.' But he could not overlook the reality that Fermilab was not the users' paradise he had described in 1963. He transformed and modernized the experimental program and its facilities, cultivating the idea that leadership of experiments should come from the users, not from the director.

As the experiments increased in scale and cost during the 1980s, the complexity of their management tasks multiplied. As the lead times grew longer, the flow of jobs clogged and Lederman was forced to deal with tighter schedules and running times. In an effort to approve larger-scale exeriments, Lederman and his Program Advisory Committee cut the number of experiments drastically, but the laboratory still did not have adequate resources to meet all user demands. But while Wilson responded by elevating frugality to a virtue, Lederman would pragmatically ask whether each experiment really needed all the requested resources to achieve its physics goals and then he found the funds.

The greater scale and complexity of experiments planned for Tevatron II pushed the issue of computing as never before. Wilson had been philosophically opposed to computing because it violated his sense of what it meant to be a physicist. Lederman met the needs of experimenters by supporting various initiatives, including the building of the Feynman Computing Center.[49]

Lederman also worked to broaden the research staff, expanding in theory and astrophysics and attracting many specialists to the laboratory. He hired one of the world's leading theoretical physicists, James Bjorken, to invigorate the interplay of experiment with theory. To help with detector planning, he brought in Georges Charpak, who had invented unique scintillation detector materials at CERN.

*Cultivating Democracy*

Lederman took a personal interest in building community. He was considered an approachable director who enjoyed and encouraged contact with his staff. For instance, he instituted a Director's coffee held every afternoon, where he regularly mingled with colleagues. Once a year he hosted a "Run with the Director," at which employees jogged with him around the 4-mile Main Ring. His personal touch extended to cultivating friends and allies for Fermilab by inviting agency officials and international leaders to his home on site. Lederman made every effort to boost morale by spending time with his staff and showing appreciation for their achievements, often with parties.

Decision-making had a greater semblance of democracy under Lederman than under Wilson. As illustrated in the way in which Lederman approached the decision on the Doubler, he typically began by seeking the advice of experts. For instance, to review the design of the antiproton source for Tevatron I, he called in Maury Tigner of Cornell, whose advice resulted in an improved design employing stochastic cooling. Similarly in May 1983, Lederman appointed Joseph Ballam from SLAC to head a committee to decide on Fermilab's future computing needs. Powerful computing innovations were then incorporated into the experimental program.[50] Lederman often turned to the physicists he appointed to his Senior Advisory Group (SAG), or in his first years to those on the "Underground Parameters Committee" (UPC), a group of accelerator physicists determined to discuss the design and performance of the Doubler.

Unlike Wilson, Lederman felt no need to manage details. Typically he would set a direction and delegate the achievement of a job to others while he moved on to the next task. This style of leadership allowed myriad activities to go on simultaneously. According to A. Lincoln Read, Lederman "would let a thousand flowers bloom and encourage people to stretch their own imaginations, thinking through a proposal in whatever way they saw fit."[51]

*Lederman's Cultural Legacy*

Lederman enlisted many volunteers and paid employees to institute cultural and social improvements at the laboratory and to bolster local support from neighboring communities. He added an on-site day-care center, a restaurant (Chez Leon), a users' center, and a gymnasium. He began a distinguished lecture series, increased the number of performing arts programs, and established ongoing exhibits in the laboratory's art gallery. He extended Wilson's Prairie Restoration Project by pursuing the formal designation of Fermilab as a National Environmental Research Park (NERP). He expanded Wilson's history of accelerators project into a larger history of particle physics program.

Lederman's legacy of science education outreach began in 1980 with the start of Saturday morning physics classes for local high-school students. Fellowships for talented particle physicists and minority students allowed Fermilab to compete with universities for staff. In 1982, Lederman established the Friends of Fermilab, a not-for-profit, grass-roots organization to raise support for innovative national math and science education programs, such as the Summer Institute for Science Teachers. Working with the Fox River Valley Industrial Association to support regional advances in math, science, and computer education, he enlisted the Friends of Fermilab to create the Illinois Mathematics and Science Academy, a statewide residential program in Aurora for gifted high-school students. Later Lederman and the Friends of Fermilab worked with educators to found the Teachers Academy for Math and Science in Chicago.

To nurture technology transfer from research at Fermilab to the larger world, Lederman established Fermilab's Industrial Affiliates. One example of such transfer was the Loma Linda Medical Accelerator, which was built at Fermilab and began treating cancer patients in California in the spring of 1989. Lederman served as chairman of the Governor of Illinois' Science Advisory Committee, and he was Vice Chairman of the Illinois Coalition, promoting prosperity in Illinois through investment in science and technology enterprises. Lederman also planned a scientific exchange program with developing nations in Latin America, including Argentina, Brazil, Colombia, and Mexico.[52]

## Conclusion

The creation and work of a large research laboratory requires the collaboration of hundreds, often thousands, of workers. Their projects and style of activity is shaped by the visions of their leaders. In the case of Fermilab, the creation and early decades of work at the laboratory was shaped by Wilson's founding myth, which in turn reflected his self-image as expressed through the figures of the frontiersman, the craftsman, and the Renaissance man. But the unification of science, art, and spirit that Wilson sought was constrained by the local conditions within which he worked in the late 1960s and through the 1970s.

One constraint was place. He construed his midwest prairie site as a frontier comparable to that encountered by pioneers traveling West in an earlier period of American history. The best and the brightest physicists were used to conducting research at the established institutions on the East and West coasts. To attract them to Illinois, he drew on compelling frontier imagery and rhetoric about the self-reliant individual who heroically overcomes obstacles in the wilderness, and he also appealed to the spiritual uplift attained through the unification of art, science, and Nature.

Limited funding was another constraint. To help compensate, he exploited imagery about the resourcefulness, independence, and manual craftsmanship exemplified by the blacksmith of his youth. He glorified frugality to the level of virtue. Drawing on innovative designers he succeeded in producing a workable technology that exceeded the specified 200-GeV design.

Over time, the inconsistencies of Wilson's vision angered and disappointed researchers.  Experimentalists needed funding, not virtue, to compete on the world stage.  Wilson's philosophy of frugality conflicted with the notion of a permanent laboratory and with the reliability required by experimentalists seeking to make indisputable discoveries.  The lone frontiersman performing quick and dirty experiments could not outperform large amply-funded collaborating teams performing large-scale frontier experiments.  And as the Washington climate for research patronage changed, Wilson could no longer lead his laboratory successfully.  Lederman's more pragmatic vision better matched the period in which he served as director.  He converted the laboratory that Wilson had built frugally for the short-term into a facility having a more stable scientific, cultural, and funding environment.

In his exceptional history of the rise of the modern business enterprise, *The Visible Hand*, Alfred Chandler highlighted the importance of a stablizing second phase that follows the rapid growth characteristic of the founding phase of the business.  Now the managers who come into control concern themselves with integration, coordination, efficient administration, and reorganization to create stability for the long term.[53]  Chandler lamented that the historians and economists writing on his subject before him had focused on the entrepreneurs or the financiers -- those associated with the founding of the company.  "They have paid almost no notice at all to the managers who, because they carried out a basic new economic function, continued to play a far more central role in the operations of the American economy than did the robber barons, industrial statesmen, or financiers."  Chandler's analysis raises an important question for the study of large national laboratories:  Do such research labs, if they are to endure, *require* a phase in which the director stabilizes the rapid expansion that follows the laboratory's creation?

**Acknowledgments**

We are grateful to Catherine Westfall and Mark Bodnarczuk for their important contributions to this paper.  We also thank Roger H. Steuwer for his careful editorial work on it.

Department of History
University of Illinois
309 Gregory Hall
Urbana, IL  61801  USA
e-mail: hoddeson@uiuc.edu

Fermi National Accelerator Laboratory
P.O. Box 500
Batavia, IL 60510 USA
e-mail: adrienne@fnal.gov

[1] L. Lederman to L. Hoddeson, private communication, fall 1978.

[2] Catherine Westfall, "The First `Truly National Laboratory': The Birth of Fermilab," (Michigan State University, Ph.D. Dissertation, 1987); C. Westfall and L. Hoddeson, "Thinking Small in Big Science," *Technology and Culture* **37** (1996) 457-492; A. Kolb and L. Hoddeson, "A New Frontier in the Chicago Suburbs: Settling Fermi National Accelerator Laboratory, 1963-1972," *Illinois Historical Journal* **88** (1995), 2-18.

[3] L. M. Lederman, "Wilson and Fermilab," in *Aesthetics and Science: Proceedings in Honor of Robert R. Wilson*, (Batavia: Fermilab, 1979), pp. 1-23.

[4] L. Hoddeson, "The First Large-Scale Application of Superconductivity: The Fermilab Energy Doubler, 1972-1983," *Historical Studies in the Physical and Biological Sciences* **18** (1987), 25-54.

[5] L. Lederman, "The State of the Laboratory," *Fermilab 1988: Annual Report of the Fermilab National Accelerator Laboratory*, pp. 1-14, on p. 4. All documents cited here that have not been attributed can be found in the Fermilab History Collection, Batavia, Illinois.

[6] R. R. Wilson, "Particles, Accelerators and Society," 1968 Richtmyer Lecture, *American Journal of Physics* **36**, (1968), 490-95, quote on 494.

[7] Frederick Jackson Turner, "The Significance of the Frontier in American History," speech delivered in 1893, copyright 1920, reprinted in *The Turner Thesis: Concerning the Role of the Frontier in American History*, ed. George Rogers Taylor (Boston: D. C. Heath and Co., 1956), pp. 1-18; See also Turner, "Contributions of the West to American Democracy" (1903), *ibid.*, pp. 19-33; William Cronon, "Revisiting the Vanishing Frontier: The Legacy of Frederick Jackson Turner," *The Western Historical Quarterly*, (April 1987), 157-76; and Richard White, "Frederick Jackson Turner," *Historians of the American Frontier*, ed. John R. Wunder (New York: Greenwood, 1988), pp. 660-681.

[8] Vannevar Bush, *Science, the Endless Frontier* (1945) (Washington: NSF, 1960). Kolb and Hoddeson, "A New Frontier in the Chicago Suburbs" (ref. 2).

[9] J. L. Heilbron and Robert W. Seidel, *Lawrence and His Laboratory: A History of the Lawrence Berkeley Laboratory* (Berkeley: University of California Press, 1989); L. Hoddeson, P. Henriksen, R. Meade, and C. Westfall, *Critical Assembly: a History of Los Alamos During the Oppenheimer Years, 1943-1945* (New York: Cambridge University Press, 1993).

[10] Rebecca Lowen, *Creating the Cold War University: The Transformation of Stanford* (Berkeley: University of California Press, 1997); Robert P. Crease, *Making Physics: A Biography of Brookhaven National Laboratory, 1946-72* (Chicago: University of Chicago Press, 1999).

[11] Philip J. Hilts, "The Main Ring," in *Scientific Temperaments* (New York: Simon and Schuster, 1982), p. 18.

[12] R. R. Wilson, interview by Hoddeson, January 12, 1979.

[13] R. R. Wilson, "Particles, Accelerators and Society," op cit., note 6.

[14] *Atomic Energy Commission Authorizing Legislation, Fiscal Year 1970, Part 1,* "Hearings Before the Joint Committee on Atomic Energy," Congress of the United States, Ninety-first Congress, April 17-18, 1969, pp. 112-118, quote on p. 113.

[15] Robert Rathbun Wilson, "From Frontiersman to Physicist" (based on an interview with Wilson by Spencer Weart, May 19, 1977) *Physics in Perspective* **2** (2000), 141-203.

[16] Silvan Schweber, "The Empiricist Temper Regnant: Theoretical Physics in the United States, 1920-1950," *Historical Studies in the Physical and Biological Sciences* **17** (1986), 55-97; and Hoddeson, Henriksen, Meade, and Westfall, *Critical Assembly* (ref. 9), chapter 1.

[17] See for example, Allen G. Debus, *Man and Nature in the Renaissance* (Cambridge: Cambridge University Press, 1978), pp. 1-15.

[18] R. R. Wilson, "Sanctimonious Memo # 137," July 11, 1969.

[19] Robert R. Wilson to Edwin McMillan, September 27, 1965; Robert R. Wilson, "Some Proton Synchrotrons, 100-1000 GeV," September, 1965.

[20] L. Hoddeson, A. Kolb, and C. Westfall, *Frontier Rings: Big Science and Particle Physics at Fermilab: 1965-89*, manuscript in progress; Kolb and Hoddeson, "A New Frontier" (ref. 2); Westfall and Hoddeson, "Thinking Small" (ref. 2); Westfall, "The First `Truly National Laboratory': The Birth of Fermilab" (ref. 2), pp. 9-200; Hoddeson, "Establishing KEK in Japan and Fermilab in the U.S.: Internationalism and Nationalism in High Energy Accelerator Physics During the 1960s," *Social Studies of Science*, **13**(1983), 1-48.

[21] Norman Ramsey to Robert R. Wilson, February 6, 1967. Also Hoddeson interview with Norman Ramsey, February 26-27, 1980.

[22] Leon Lederman, "The Truly National Laboratory," in *1963 Super-High-Energy Summer Study*, Brookhaven National Laboratory, AADD-6.

[23] Lederman, *ibid.*, p. 9. Also see Westfall, "The First `Truly National Laboratory'" (ref 2), pp. 9-200.

[24] John Peoples, E. Goldwasser, private communications, May 25, 2001.

[25] *National Accelerator Laboratory: Design Report, January 1968*, Universities Research Association, under the auspices of the United States Atomic Energy Commission, pp. 3-11.

[26] Wilson, "Frontiersman to Physicist" (ref. 15).

[27] "Edited Version of Talk at 1st NAL Users Meeting," April 7, 1967. Also see Hoddeson interview with M. Hankerson, September 14, 1978.

[28] Robert R. Wilson, "Wilson to Whole Staff," October 1, 1969; Robert R. Wilson, "Initial Experiments at NAL," January 25, 1973, presented at the Coral Gables Conference.

[29] Robert R. Wilson and Adrienne Kolb, "Building Fermilab: A User's Paradise," in L. Hoddeson, L. M. Brown, M. Riordan, and M. Dresden, eds., *The Rise of the Standard Model: Particle Physics in the 1960s and 70s*, (New York: Cambridge University Press, 1997), pp. 338-363.

[30] Robert R. Wilson to Edwin McMillan, September 27, 1965, Robert R. Wilson, "Some Proton Synchrotrons, 100-1000 GeV," September 1965.

[31] Trevor Pinch, *Confronting Nature: The Sociology of Solar Neutrino Detection* (Dordrecht: Reidel, 1986); Robert Smith, *The Space Telescope: A Study of NASA, Science, Technology, and Politics* (Cambridge: Cambridge University Press, 1989).

[32] Robert R. Wilson, "My Fight Against Team Research," *Daedalus* **99**, No.. 4., (Fall 1970), 1076-1087.

[33] For a broader discussion of the rift between large laboratories and government see Westfall, "Science Policy and the Social Structure of Big Laboratories," in Hoddeson, Brown, Riordan, and Dresden, *Rise of the Standard Model* (ref. 30), pp. 364-83; also Chrisman interview by L. Hoddeson and A. Kolb, February 22, 2001.

[34] *FermiNews* **1**, No. 25 (October 26, 1978) 1.

[35] Leon Lederman with Dick Teresi, *The God Particle: If the Universe is the Answer, What is the Question?* (New York: Houghton Mifflin, 1993), p. 7.

[36] Malcolm W. Browne, "Leon Lederman: Captain of Science," *Discovery* (October 1981), 45-50, on 48.

[37] Lederman with Teresi, *God Particle* (ref. 36), p. 8.

[38] E. L. Goldwasser, "Highlights of the Summer PAC Meeting," *NALREP* (July 1976), pp. 1-11; interview with Alvin Tollestrup by L. Hoddeson and K. Paik, FNAL, July 1990.

[39] L. M. Lederman to scientific staff, October 26, 1978.

[40] L. Hoddeson, "The First Large-Scale Application of Superconductivity: The Fermilab Energy Doubler, 1972-1983," *Historical Studies in the Physical and Biological Sciences* **18** (1987), 25-54.

[41] *Ibid.*

[42] "Leon M. Lederman," *Current Biography* **50**, No. 9 (September 1989), 17-21, quote on 20.

[43] See Robert Crease's study of ISABELLE, in progress.

[44] Chrisman interview by Hoddeson and Kolb (ref. 34).

[45] Interview with A. Mravca by L. Hoddeson, December 14, 1983.

[46] Lederman private communication. to A. Kolb, June 2001.

[47] Peter Galison, *Image and Logic* (Chicago: University of Chicago Press, 1997). Also Hoddeson, Kolb, and Westfall, "The Transition to Megascience: Colliding Beams at Fermilab," in *Frontier Rings* (ref. 22), chapter 11.

[48] M. Riordan, L. Hoddeson, R. Jacobs, A. Kolb, G. Sandiford, and S. Weiss, *Tunnel Visions: the Rise and Fall of the Superconducting Super Collider*, manuscript in progress. Also see Michael Riordan, "The Demise of the Superconducting Super Collider," *Physics in Perspective* **2** (2000), 411-425.

[49] Kevin A. Brown, "Feynman Center Opens New Era for Fermilab Computing," *FermiNews* **XII,** No. 4., (March 10, 1989), pp. 1-2, 5.

[50] Jeffrey A. Appel, "The Computing Department," *Fermilab 1988: Annual Report of the Fermi National Accelerator Laboratory* (Batavia, Fermilab, 1989), 75-79.

[51] Private communication to M. Bodnarczuk by A. Lincoln Read.

[52] *Ibid.*

[53] Alfred D. Chandler, Jr., *The Visible Hand* (Cambridge, Mass: Harvard University press, 1977), pp. 160-76, 490-497. Thomas S. Hughes, *American Genesis: A Century of Invention and Technological Enthusiasm,* 1870-1970 (New York: Viking, 1989), pp. 226-46. Similar stages are detailed by Thomas P. Hughes in *Networks of Power: Electrification in Western Society (1800-1930)* (Baltimore: Johns Hopkins University Press, 1983).